\documentclass[twocolumn,preprintnumbers,amsmath,amssymb,amssymb,aps, showpacs,prl]{revtex4-1}
\usepackage{graphicx}
\usepackage{dcolumn}
\usepackage{bm}
\usepackage{braket}
\begin{document}
\title{Analytical solutions for a two-level system driven by a class of chirped pulses.}
\author{Pankaj K. Jha$^{1,*}$}
\author{Yuri V. Rostovtsev$^{1,2}$}%
\affiliation{$^{1}$Institute for Quantum Science and Engineering and
 Department of Physics, Texas A\&M University, TX 77843.\\
$^{2}$Department of Physics, University of North Texas,1155 Union Circle  311427, Denton, TX 76203}
\date{\today}
\begin{abstract}
We present analytical solutions for the problem of a two-level atom driven by a class of chirped pulses. The solutions are given in terms of Heun functions. Using appropriate chirping parameters an enhancement of four-orders of magnitudes in the population transfer is obtained.
\end{abstract}
\maketitle
Interaction of coherent optical pulses with quantum systems is a fundamental problem~\cite{fastcars,A1}. Nowadays, the laser systems produce controlled intense ultra-short optical pulses~\cite{lasers}. Various technologies have been used for pulse shaping~\cite{pulse-shaping} that allows researchers to provide coherent optical control \cite{coherent-control} of excitation in quantum systems that has a broad range of applications from nonlinear laser spectroscopy to generation of coherent radiation. For example, the chirped pulses~\cite{pulse-shaping,fastcars} are used to produce maximal coherence in atomic and molecular systems.
Maximal coherence can be used for generation of short-wavelength of radiation that has been a focus of research recently where an atomic system under the action of a far-off resonance strong
pulse of laser radiation has been considered and it has been shown that such pulses can excite remarkable coherence on high frequency far-detuned transitions; and this coherence can be used for effcient generation of soft x-ray and ultraviolet (XUV) radiations~\cite{A1,A2,A3}.\,Also maximal coherence can be used for molecular spectroscopy, for example, time-resolved coherent Raman spectroscopy, to obtain molecule-specific signals from molecules, which can serve as a marker molecule for bacterial spores~\cite{fastcars}.

In this brief report we will investigate two class of chirped pulses for which the problem can be solved exactly in analytical form.\,Using the appropriate chirping parameters, the population transfer, after the the pulse is gone, can be optimized and for the pulse considered here, four-order of magnitudes enhancement was obtained.\,Unchirped pulse corresponding to Heun and Confluent Heun equation has been recently investigated extensively~\cite{A3} where we have included an estimate of energy of emission of soft x-ray and ultraviolet radiation via excited quantum coherence in the atomic system. The estimate shows good potential for a source of coherent radiation based on the discussed mechanism.

The equation of motion for the probability amplitudes for the states $|a\rangle$ and $|b\rangle$ (see Fig\,1a) of a two-level atom (TLA)~\cite{B1,B2} interacting with a classical field (under rotating-wave approximation RWA) with non-zero chirping~\cite{C1}. is given as
\begin{figure}[t]
 \includegraphics[height=3.8cm,width=0.20\textwidth,angle=0]{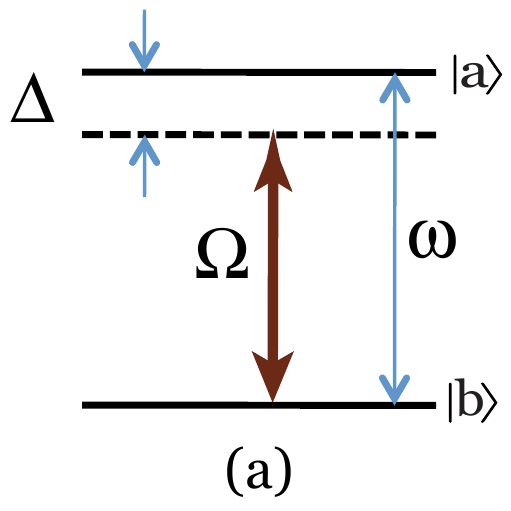}
  \includegraphics[height=3.5cm,width=0.26\textwidth,angle=0]{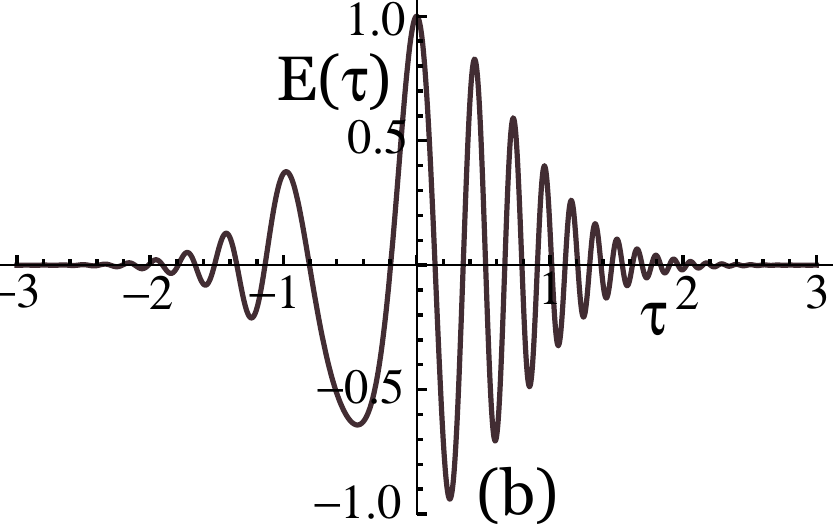}
  \caption{(Color online) (a)Two-level atomic system, atomic transition frequency $\omega=\omega_{a}-\omega_{b}$, detuning $ \Delta = \omega -\nu$ and Rabi frequency $\Omega(t)=\wp{\cal E}(t)/2\hbar$. (b) Classical electromagnetic field E(t)$=\mbox{exp}(-\alpha^2 t^2)\mbox{cos}(\nu t + t^{2})$}
\end{figure}
\begin{subequations}\label{I3}
\begin{align}
\dot{C}_{a}&= i\Omega(t) e^{i\vartheta(t)}C_{b},\label{second}\\
\dot{C}_{b}&= i\Omega^{*}(t)e^{-i\vartheta(t)}C_{a},
\end{align}
\end{subequations}
\noindent where $\vartheta(t)=\Delta t +\phi(t)$. Here $\Delta=\omega-\nu$~\cite{P} and $\Omega(t)=\wp\mathcal{E}(t)/2\hbar$. Let us define the dimensionless parameters as 
\begin{equation}\label{I4}
\tau = \alpha t, \quad \beta = \frac{\Delta}{\alpha}, \quad \gamma = \frac{\Omega_{0}}{\alpha}.
\end{equation}
To solve for $C_{a}$, we can get a second order linear differential equation for $C_{a}(t)$ from Eq.(\ref{I3}), which in terms of the dimensionless parameters Eq.(\ref{I4}) is given as
\begin{equation}\label{I5}
\ddot{C}_{a}-\left[i\beta + \frac{\dot{\Omega}(\tau)}{\Omega(\tau)}+i\dot{\phi}(\tau) \right]\dot{C}_{a}+\Omega^{2}(\tau)C_{a}=0.
\end{equation}
\subsection{Class I: Heun Equation}
To find analytical solution for Eq.(\ref{I5}), We introduce a new variable $ \varphi=\varphi(\tau)$ defined by
\begin{equation}\label{I6}
\tau =(1/2)\mbox{ln}[\varphi^{\mu}/(1-\varphi)^{\mu+\lambda}],
\end{equation}
and make an ansatz for the pulse envelope $\Omega(\tau)$ and the chirping function $\phi(\tau)$ as
\begin{subequations}\label{I7}
\begin{align}
\Omega(\tau)&=\left[\frac{2\varphi(1-\varphi)}{(c-\varphi)}\right]^{1/2}\left(\frac{\gamma}{\mu+\lambda\varphi}\right),\\
\dot{\phi}(\tau)&=\left\{\frac{-2c\,\zeta+2[(\zeta+\xi)+c(\zeta + \eta)]\varphi}{(\varphi-c)(\mu +\lambda \varphi)}\right\}.
\end{align}
\end{subequations}
\begin{figure*}[t]
  \includegraphics[height=3.6cm,width=0.98\textwidth,angle=0]{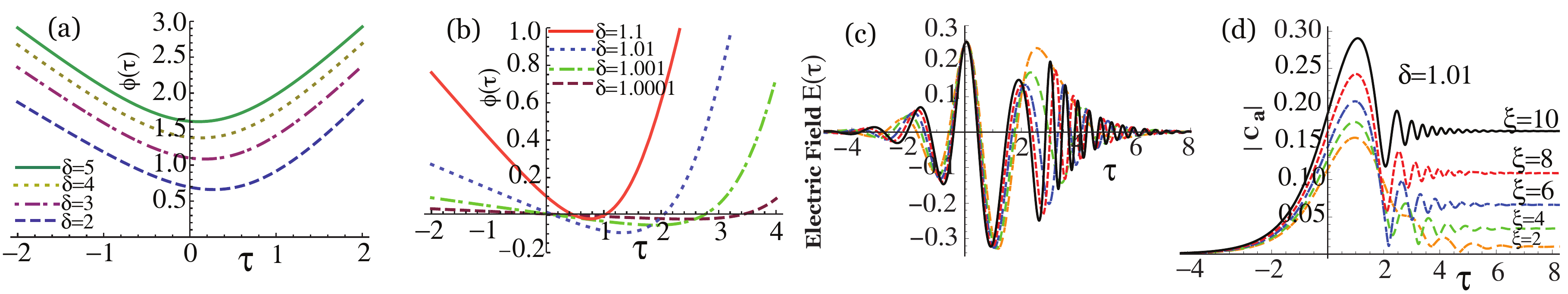}
  \caption{(Color online)(I) Heun Equation case: Chirping function $\phi(\tau)$ given by Eq.(\ref{I20}) for $\xi=10$ and (a) $\delta >1$,\,(b) $\delta \approx$ 1 (c) The Electric field E$(\tau)$ for varying $\xi$ and $\delta=1.01$. (d) Probability amplitudes for the upper level $|a\rangle$ for the corresponding fields in (c). $|C_{a}(\tau)|$ is given by Eq.(\ref{I23})}
\end{figure*}
In terms of the variable $\varphi(\tau)$ and the definition of $\Omega(\tau), \dot{\phi}(\tau)$ from Eq.(\ref{I7}), Eq.(\ref{I5}) takes the form
\begin{equation}\label{I8}
C^{''}_{a}+\left[\frac{\rho}{\varphi}+\frac{\sigma}{\varphi-1}+\frac{\upsilon}{\varphi-c}\right]C^{'}_{a}+\frac{ab\varphi-q}{\varphi (\varphi-1)(\varphi-c)}C_{a}=0,
\end{equation}
where $(c>1)$ and
\begin{equation}\label{I9}
\begin{split}
\rho&= \frac{1}{2}-i\left(\zeta+\frac{\beta\mu}{2}\right),\,\sigma= \frac{1}{2}+i\left[\frac{\beta(\mu +\lambda)}{2}-\eta\right],\\
\upsilon&= \frac{1}{2}-i\xi,\, q=-\frac{\gamma^{2}}{2},\, a= 0,\, b=\frac{1}{2}+\frac{i\beta \lambda}{2},\, c=\frac{\delta +1 }{2}.
\end{split}
\end{equation}
The parameters of a Heun Equation~\cite{B3,HS} are constrained, by the general theory of Fuschsian equations as,
$\rho+\sigma+\upsilon=a+b+1$ which provides us a the first constraint relation for the chirping parameters $\zeta, \eta, \xi$ as 
\begin{equation}\label{I11}
\zeta+\eta+\xi=0.
\end{equation}
The quantity $\nu +\dot{\phi}$ is the instantaneous pulse frequency; thus $\dot{\phi}$ should vanish for maximum of $\Omega(\tau)$. From Eq.(\ref{I7}a), we get the corresponding $\varphi_{0}$ which satisfy the equation
\begin{equation}\label{I12}
\lambda \varphi^{3}-(2\lambda+\mu)\varphi^{2}+c(\lambda+2\mu)\varphi-c\mu=0.
\end{equation} 
Thus the second constraint relation for the chirping parameters is given as 
\begin{equation}\label{I13}
\left\{\frac{-2c\,\zeta+2[(\zeta+\xi)+c(\zeta + \eta)]\varphi_{0}}{(\varphi_{0}-c)(\mu +\lambda \varphi_{0})}\right\}=0.
\end{equation} 
The general solution for Eq.(\ref{I8}), which has regular singularity at $\varphi=0$ is given in terms of Heun local solutions, ${\cal{H}}l(\varphi)$ as,
\begin{equation}\label{I14}
  \begin{split}
   & C_{a}={\cal{P}}_{1}\varphi^{1-\rho}{\cal{H}}l[c,q+(1-\rho)((c-1)\sigma+a+b-\rho+1);\\
    & a-\rho+1,b-\rho+1,2-\rho,\sigma;\varphi]+{\cal{P}}_{2}\,{\cal{H}}l\left[c,q;a,b,\rho,\sigma;\varphi\right],
  \end{split}
\end{equation}
where the constants, ${\cal{P}}_{1} ,{\cal{P}}_{2}$ can be found using the initial conditions of the system. In the limit $\tau \rightarrow \infty $, the population left in the level $|a\rangle$ can be obtained by substituting $\varphi \rightarrow 1$ in Eq.(\ref{I14}). Let us consider a simple case of $\mu=1, \lambda=0$ in Eq.(\ref{I6}) and Eq.(\ref{I7}a) gives
\begin{subequations}\label{I16}
\begin{align}
\varphi(\tau)&=\frac{1+\mbox{tanh}(\tau)}{2}, \label{second} \\
\Omega(\tau)&=\gamma \left[\frac{2\varphi(1-\varphi)}{(c-\varphi)}\right]^{1/2}.
\end{align}
\end{subequations}
From Eq(\ref{I16}) the pulse takes the form
\begin{equation}\label{I17}
\Omega(\tau)=\frac{\gamma\, \mbox{sech}(\tau)}{\sqrt{\delta -\mbox{tanh}(\tau)}}, \quad \delta =2c-1.
\end{equation}
This pulse shape serves as an excellent model for a smooth box pulse, by taking care of non-analyticity at its edges with the help of the pulse parameter $\delta$. Using Eq.(\ref{I12}) we get $
\varphi_{0}=\,c\pm \sqrt{c^{2}-c}$. From one of our earlier assumptions $c>1$ only one of the possible values is allowed for $\varphi_{0}$ as $0\le \varphi_{0}\le 1$. Subsequently using $\varphi_{0}=c-\sqrt{c^{2}-c}$ in Eq.(\ref{I13}), we get the constraint equation as
\begin{equation}\label{I19}
\frac{\zeta-\eta}{\xi}=\delta-\sqrt{\delta^{2}-1}.
\end{equation}
The defining equation for the chirping function takes the form
\begin{equation}\label{I20}
\phi(\tau)=\xi\left\{\left(\delta-\sqrt{\delta^{2}-1}\right)\tau +\,\text{ln}[\delta\,\text{cosh}(\tau)-\text{sinh}(\tau)]\right\}.
\end{equation}
\noindent For the pulse defined by Eq.(\ref{I17}),
using the scaling parameters Eq.(\ref{I4}) and the chirping function Eq.(\ref{I20}), Eq.(\ref{I5}) gives
\begin{align}\label{I21}
  \begin{split}
    &\ddot{C}_{a}(\tau)-\left[\frac{1}{2}\left(\frac{1-2\delta \mbox{tanh}\tau + \mbox{tanh}^{2}\tau}{\delta - \mbox{tanh}\tau}\right)-i\xi\left(\sqrt{\delta^{2}-1}\right.\right.\\
   &\left.\left.-\frac{\delta^{2}-1}{\delta -\text{tanh}(\tau)}\right)+i\beta \right]\dot{C}_{a}(\tau)
    +\frac{\gamma^{2}\mbox{sech}^{2}\tau }{\delta-\mbox{tanh}\tau}C_{a}(\tau)=0.
  \end{split}
\end{align}
Let us define the initial conditions for our system as
\begin{equation}\label{I22}
C_{a}(\tau \rightarrow -\infty)=0, \quad |C_{b}(\tau \rightarrow -\infty)|=1.
\end{equation}
Solution for Eq.(\ref{I21}), satisfying the initial conditions is give as 
\begin{equation}\label{I23}
\begin{split}
C_{a}={\cal{P}}&\varphi^{1-\rho}{\cal{H}}l[c, q+(1-\rho)((c-1)\sigma+a+\\
&b-\rho+1);a-\rho+1,b-\rho+1,2-\rho,\sigma;\varphi],
\end{split}
\end{equation}
where $\varphi(\tau)$ is given by Eq.(\ref{I16}a) and the Heun parameters as . 
\begin{equation}\label{I24}
\begin{split}
&\rho= \frac{1}{2}-i\left(\zeta+\frac{\beta}{2}\right),\,\sigma= \frac{1}{2}+i\left[\frac{\beta}{2}-\eta\right],\, \\
\upsilon&= \frac{1}{2}-i\xi,\,q=-\frac{\gamma^{2}}{2},\, a= 0,\, b=\frac{1}{2},\, c=\frac{\delta +1 }{2}.
\end{split}
\end{equation}
The chirping parameters $\zeta, \eta$ and the constant ${\cal{P}}$ are given as 
\begin{subequations}\label{I25}
\begin{align}
\zeta& = -\frac{\xi}{2}\left(1-\delta + \sqrt{\delta^{2}-1}\right), \\
\eta &= -\frac{\xi}{2}\left(1+\delta - \sqrt{\delta^{2}-1}\right), \\
{\cal{P}} &=i\gamma  \left[\frac{2^{(1-i\xi)}(1+\delta)^{i\xi -1/2}}{1+i(2\zeta+\beta)}\right].
\end{align}
\end{subequations}
Here we have kept $\xi$ as a free parameter for the chirping function $\phi(t)$. In Fig(2), we have considered some forms of the chirping function $\phi(\tau)$ [see Fig(2a,2b)] for $\delta > 1$ and $\delta \approx 1$ respectively, given by Eq.(\ref{I20}). Influence of chirping on the evolution of the probability amplitude for the upper level $|a\rangle$ in shown in Fig(2d) for the corresponding pulses in Fig(2c)
\subsection{Class II: Confluent Heun Equation}
In this section we will consider another class of pulse and the corresponding chirping function. Let us define the pulse and the chirping function as 
\begin{subequations}\label{II26}
\begin{align}
\Omega(\tau)&=\frac{2\sqrt{2}\,\gamma(1-\varphi)\sqrt{\varphi}}{\mu+\lambda \varphi}, \label{second}\\
\dot{\phi}(\tau)&= \frac{2 \zeta -2 (\zeta +\eta ) \varphi }{\mu +\lambda  \varphi }.
\end{align}
\end{subequations}
In terms of the variable $\varphi(\tau)$ and the definition of $\Omega(\tau), \dot{\phi}(\tau)$ from Eq.(\ref{I7}), Eq.(\ref{I5}) takes the form
\begin{equation}\label{II27}
\frac{d^{2}C_{a}}{d\varphi^{2}}+\left(\frac{u}{\varphi}+\frac{v}{\varphi-1}\right)\frac{dC_{a}}{d\varphi}+
\frac{(p\varphi+q)C_{a}}{\varphi(\varphi-1)}=0.
\end{equation}
where
\begin{equation}\label{II28}
u= \frac{1}{2}-i\left[\zeta+\frac{\beta\mu}{2}\right],\,v= i\left[\frac{\beta(\lambda +\mu)}{2}-\eta\right],\,p=-q=2\gamma^{2}.
\end{equation}
The critical point which corresponds to the peak of $\Omega(\tau)$ is given by
\begin{equation}\label{II29}
\varphi_{0}=-\left[\frac{\lambda+3 \mu - \sqrt{(\lambda +\mu ) (\lambda +9 \mu )}}{2 \lambda }\right],
\end{equation}
and the corresponding $\tau_{0}$ can be found using Eq.(\ref{I6}). At this point $\dot{\phi}=0$ which gives a constraint relation as
 \begin{equation}\label{II30}
 \begin{split}
 3 \zeta  (\lambda +\mu )-\zeta  \sqrt{(\lambda +\mu ) (\lambda +9 \mu )}+\\
 \eta  \left(\lambda +3 \mu -\sqrt{(\lambda +\mu ) (\lambda +9 \mu )}\right)=0.
 \end{split}
 \end{equation}
 The general solution of the Confluent Heun Equation Eq.(\ref{II27}) is given as
\begin{align}\label{II31}
  \begin{split}
    &C_{a}={\cal{P}}_{1} {\cal{H}}l^{(c)}[0,u-1,v-1,p,q+(1-uv)/2,\varphi] \\
    &+ {\cal{P}}_{2}\varphi^{1-u}{\cal{H}}l^{(c)}[0,1-u,v-1,p,q+(1-uv)/2,\varphi],
  \end{split}
\end{align}
where ${\cal{P}}_{1}$, ${\cal{P}}_{2}$ can be found using the initial condition of the system. In the limit $\tau \rightarrow \infty $, the population left in the level $|a\rangle$ can be obtained by substituting $\varphi \rightarrow 1$ in Eq.(\ref{II31}).
 \begin{figure}[b]
  \includegraphics[height=3.8cm,width=0.46\textwidth,angle=0]{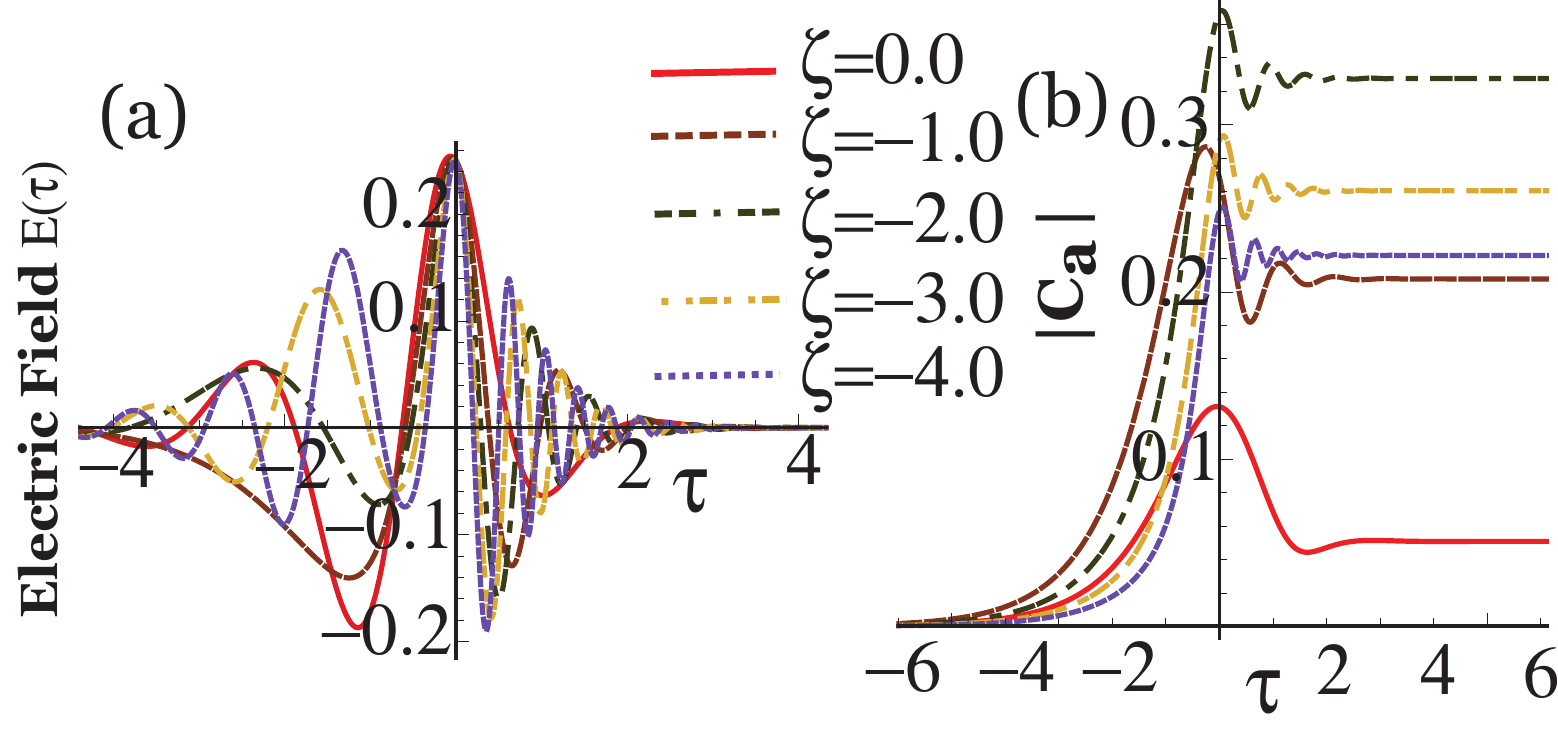}
  \caption{(Color online) (II) Confluent Heun Equation case:  (a) Profile of the Electric field E$(\tau)$ for varying $\zeta$. (b) Probability amplitudes for the upper level $|a\rangle$ for the corresponding fields in (a). The pulse envelope $\Omega(\tau)$ and $\dot{\phi}(\tau)$ is given by Eq.(\ref{II33}). }
\end{figure}
Let us consider a simple case of $\mu=1, \lambda=0$. Thus the new variable is given by Eq.(\ref{I16}a). From Eqs.(\ref{II29}, \ref{II30}) we get $\varphi_{0}=1/3$ and $\eta=2\zeta$. Pulse shape $\Omega(\tau)$ and the chirping function can be written as 
\begin{subequations}\label{II33}
\begin{align}
\Omega(\tau)&=\gamma\,\mbox{sech}(\tau)\left[1-\mbox{tanh}(\tau)\right]^{1/2}, \label{second}\\
\dot{\phi}(\tau)&= -\zeta  \left[1+3 \text{tanh}(\tau)\right].
\end{align}
\end{subequations}
\noindent For the pulse defined by Eq.(\ref{II33}a),
using the scaling parameters Eq.(\ref{I4}) and the chirping function Eq.(\ref{II33}b), Eq.(\ref{I5}) gives
\begin{align}\label{II34}
  \begin{split}
    \ddot{C}_{a}(\tau)-&\left\{i\beta -\frac{1}{2}\left[1+3\mbox{tanh}(\tau)\right]-i\zeta  \left[1+3 \text{tanh}(\tau)\right]\right\}\dot{C}_{a}(\tau) \\
    & +\gamma^{2}\mbox{sech}^{2}(\tau) \left[1-\mbox{tanh}(\tau)\right] C_{a}(\tau)=0.
  \end{split}
\end{align}
 \begin{figure}[t]
  \includegraphics[height=6.8cm,width=.49\textwidth,angle=0]{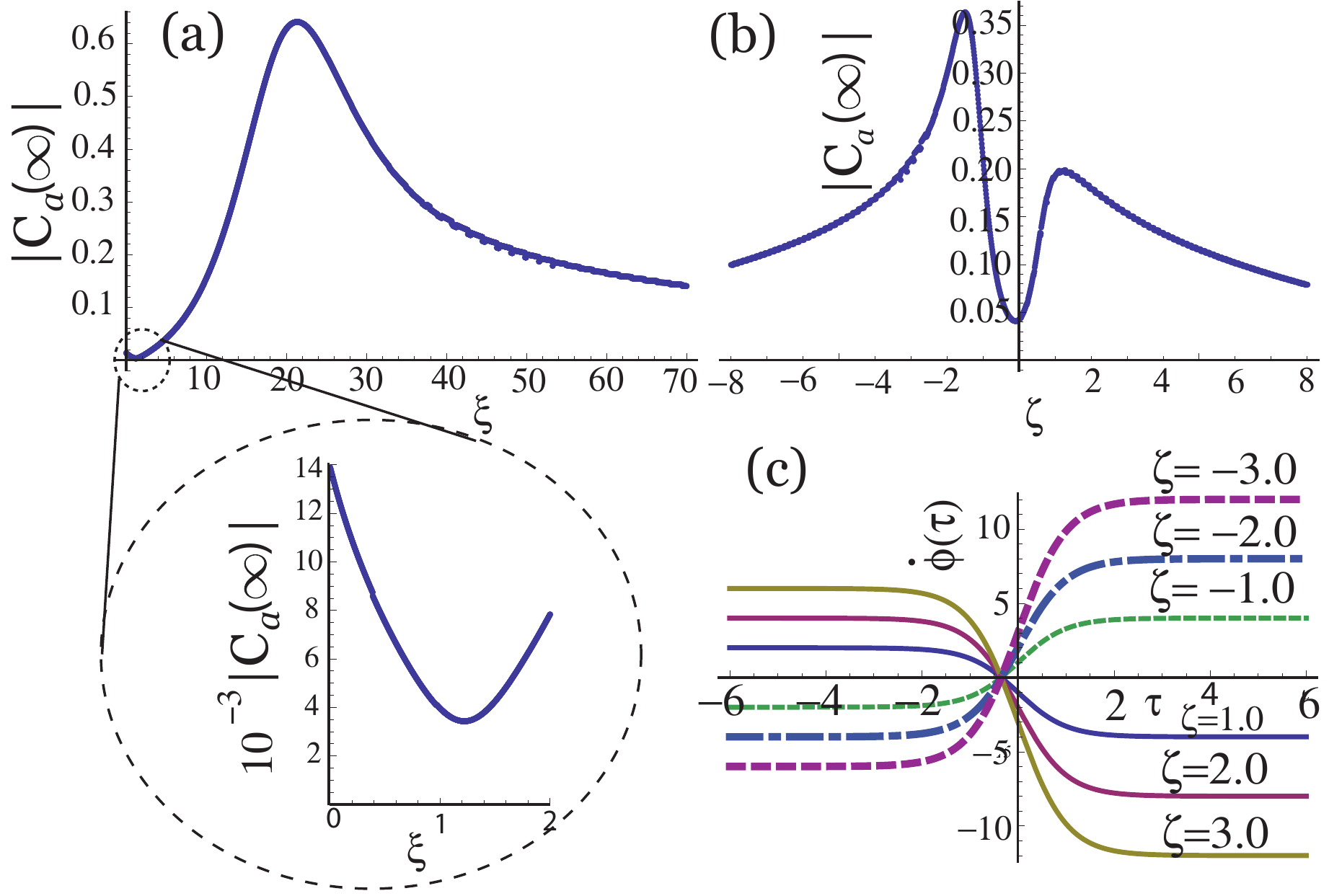}
  \caption{(Color online) Effect of chirping on the population left in the excited states $|a\rangle$. (a) Heun Equation and (b) Confluent Heun Equation. The inset shows the dip (minima) in the population left for the Heun case. For calculations $\beta =2.5, \gamma=0.25, \delta =1.01$ (c) Chirping function for the Confluent Heun case $\dot{\phi}(\tau)$}
\end{figure}
 \begin{figure}[t]
  \includegraphics[height=9.2cm,width=.48\textwidth,angle=0]{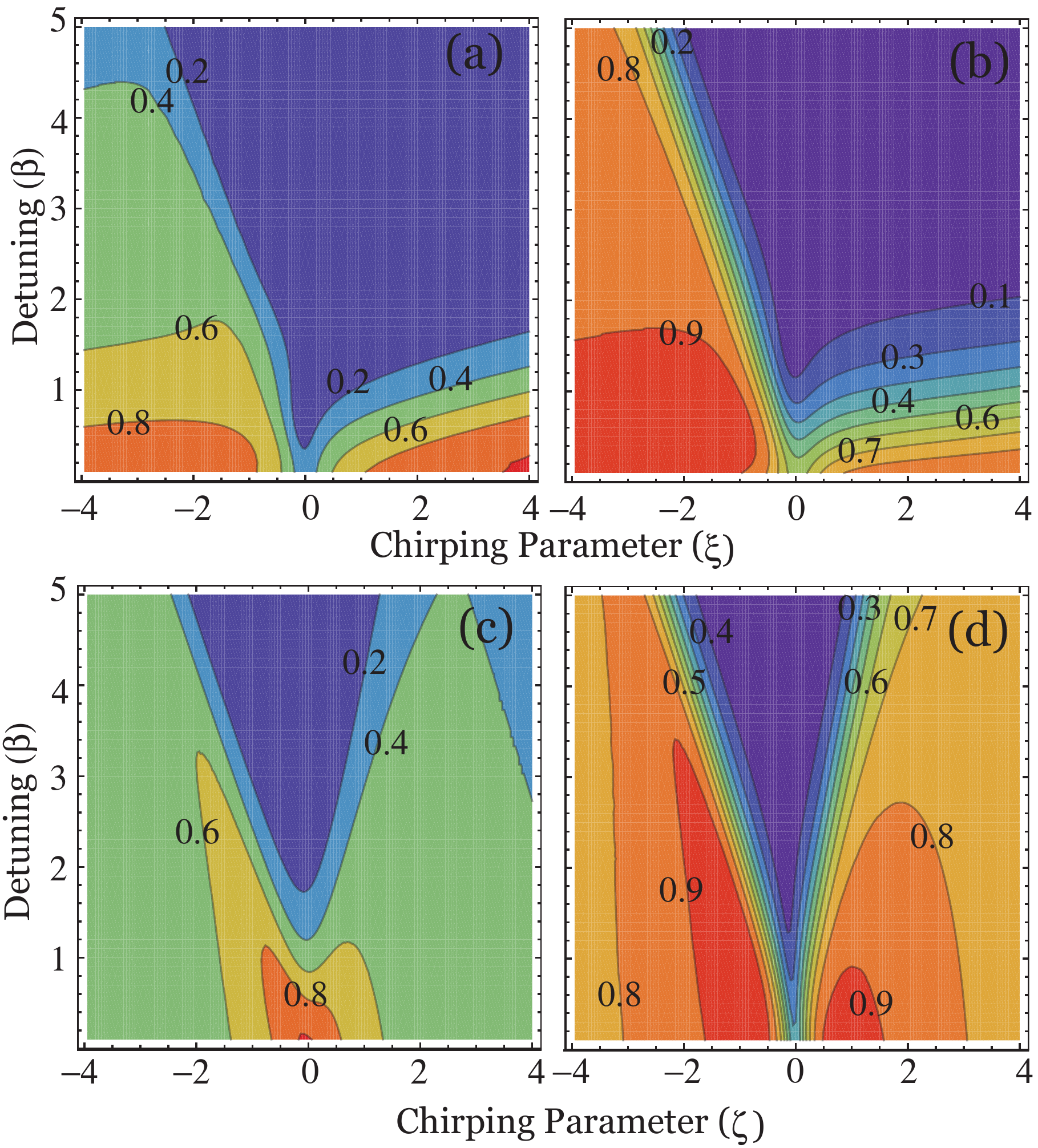}
  \caption{(Color online) Contour plot showing the effect of chirping and the detuning on the population left in the excited states $|a\rangle$ .(a) and (b) corresponds to Heun while (c) and (d) to Confluent Heun case. For calculations (a) $ \gamma=0.50, \delta =1.01$ (b) $\gamma=1, \delta=1.01$ (c) $\gamma= 0.5$ (d) $\gamma=1.$}
\end{figure}
Let us define the initial conditions for our system as Eq.(\ref{I22}). Solution for Eq.(\ref{II34}), satisfying the initial conditions is give as 
\begin{equation}\label{II35}
C_{a}= {\cal{P}}\varphi^{1-u}{\cal{H}}l^{(c)}[0,1-u,v-1,p,q+(1-uv)/2,\varphi],
\end{equation}
where $\varphi(\tau)$ is given by Eq.(\ref{I16}a) and the Heun parameters as . 
\begin{equation}\label{II36}
u= \frac{1}{2}-i\left[\zeta+\frac{\beta}{2}\right],\,v= i\left[\frac{\beta}{2}-\eta\right],\,p=-q=2\gamma^{2}.
\end{equation}
The constant ${\cal{P}}$ is given as 
\begin{equation}\label{II37}
{\cal{P}}=\gamma\left[\frac{2^{(3/2 +3i\zeta)}}{(2\zeta +\beta)-i}\right].
\end{equation}
Here we have kept $\zeta$ as a free parameter for the chirping function $\phi(\tau)$ given by Eq.(\ref{II33}). Influence of chirping on the evolution of the probability amplitude for the upper level $|a\rangle$ in shown in Fig(3b) for the corresponding pulses in Fig(3a). To see the effect of chirping on the population left in the upper level $|a\rangle$, we have plotted in Fig(4), $|C_{a}(\infty)|$ as a function of the free chirping parameter for the Heun and the Confluent Heun case for a particular choice of the detuning $\beta$ and the peak Rabi frequency $\gamma$. We see that $|C_{a}(\infty)|$ ranges from $4\cdot 10^{-3} \backsim 6\cdot 10^{-1}$. In Fig(5) we have plotted population left in the upper level $|a\rangle$ as a function of detuning $\beta$ and the free chirping parameters $\xi,\zeta$  for the Heun and the Confluent Heun case respectively.

In conclusion, we have found new solutions for the problem of two-level system driven by chirped pulses in exact analytical form. The solutions are given in terms of Heun function $\mathcal{H}l$ and the Confluent Heun function $\mathcal{H}l^{(c)}$. Using the appropriate chirping parameters we have shown that the population left after the pulse is gone can be enhanced by four-orders of magnitude [see Fig(4(a) inset]. Application of chirped pulses on the population transfer or generation of coherence makes it very interesting to look for new shapes for which the system can be solved analytically because finding exact analytical solutions for such a problem will not only supplement numerical simulations but will also be useful in understanding the underlying physics.

We thank M.O.Scully, L.Keldysh, M.S.Zubairy  for useful discussions and gratefully acknowledge the support from the NSF Grant EEC-0540832 (MIRTHE ERC), Office of Naval Research (N00014-09-1-0888 and N0001408-1-0948), Robert A. Welch Foundation (Award A-1261)) and the partial support from the CRDF. 
     

\begin{thebibliography}{99}
\bibitem{P} Here we use the convention that all frequencies are circular frequencies so that $\hbar\nu$ (not $h\nu$) is the photon energy.
\end{thebibliography}
\end{document}